\pdfoutput=1
\documentclass[11pt,a4paper]{article}
\usepackage{array}
\usepackage[footnotesize,bf]{caption}
\usepackage[dvips]{graphicx}
\usepackage{color} 
\usepackage{amsmath}
\usepackage{fancyhdr}
\usepackage[]{natbib}
\usepackage[LGRgreek]{mathastext}
\usepackage[yyyymmdd,hhmmss]{datetime}

\fancyheadoffset[R]{0pt}
\fancyfootoffset[R]{0pt}
\renewcommand{\headrulewidth}{0.4pt}
\renewcommand{\footrulewidth}{0.4pt}
\definecolor{orange}{rgb}{1.0,0.5,0.0}
\definecolor{aqgr}  {rgb}{0.0,1.0,0.6} 
\definecolor{viol}  {rgb}{0.8,0.6,0.8}
\definecolor{figdr} {rgb}{1.0,1.0,1.0} 
\definecolor{colnu} {rgb}{1.0,0.0,1.0} 
\definecolor{colhd} {rgb}{1.0,0.8,0.0} 

\newcolumntype{C}[1]{>{\centering\let\newline\\\arraybackslash\hspace{0pt}}m{#1}}
\newif\ifhpar

\title{\bfseries{\textsc{Is psychosis caused by defective dissociation? \\
   An Artificial Life model for schizophrenia}}}
   
\author{Alessandro Fontana} \date{}

\begin{document}
\maketitle
   
\clubpenalty=10000
\widowpenalty=10000

\begin{abstract}
Both neurobiological and environmental factors are known to play a role in the origin of schizophrenia, but no model has been proposed that accounts for both. This work presents a functional model of schizophrenia that merges psychodynamic elements with ingredients borrowed from the theory of psychological traumas, and evidences the interplay of traumatic experiences and defective mental functions in the pathogenesis of the disorder. Our model foresees that dissociation is a standard tool used by the mind to protect itself from emotional pain. In case of repeated traumas, the mind learns to adopt selective forms of dissociation to avoid pain without losing touch with external reality. We conjecture that this process is defective in schizophrenia, where dissociation is either too weak, giving rise to positive symptoms, or too strong, causing negative symptoms. 
\end{abstract}


\section{Theories on schizophrenia}  

\ifhpar \colorbox{colhd}{alife approach} \\ \fi
Mental disorders can be described at multiple levels: psychological, neurobiological, genetic. In this work we deal with another level, that we call functional, situated between the psychological and the neurobiological levels. The functional level aims to characterise the structure of the system under investigation and the function of each component, regardless of its neurobiological implementation. This is the scientific approach of Artificial Life: study life as it could be, to understand life as it is. This work presents a functional model of schizophrenia, which is used to interpret the psychological-symptomatic aspects of the disorder and to account for the neurobiological substrate.  

\ifhpar \colorbox{colhd}{genes and environment} \\ \fi
Both genetic and environmental factors, such as traumatic experiences, are known to play a role in the origin of schizophrenia, but few models have been proposed that account for both. The model we are going to propose merges psychodynamic elements with ingredients borrowed from the theory of psychological traumas, and evidences the interplay of traumatic experiences and defective mental functions in the pathogenesis of schizophrenia. A thorough characterisation of this complex condition is outside the scope of this work: our model, based on a simplification of reality, has the purpose to interpret some limited aspects of mental functioning.

\ifhpar \colorbox{colhd}{dopamine hypothesis} \\ \fi
Among theories of schizophrenia, much attention has been dedicated to the `dopamine hypothesis' \citep{howes2009dophyp}. Psychosis is hypothesised to be caused by the phenomenon of `aberrant salience' \citep{kapur2003}, determined by an excess of neurotransmitter dopamine in the limbic system. This theory is supported by the fact that amphetamines, which trigger the release of dopamine, may induce hallucinations and delusions \citep{laruell1996} and that antipsychotic drugs block the D2 dopamine receptor. However, large genetic studies were unable to identify schizophrenia-associated genes involved in dopamine transmission \citep{alexis2016}.

\ifhpar \colorbox{colhd}{glutamate hypothesis} \\ \fi
Another theory links schizophrenia to a malfunction of glutamate NMDA receptors which, unlike dopamine receptors, are distributed across the whole brain \citep{javitt2010}. This theory is based on the observation that some drugs (such as PCP) that block the NMDA receptor, produce symptoms similar to those of schizophrenia. Unlike amphetamines, such drugs would be able to reproduce the full spectrum of schizophrenia symptoms (positive, cognitive, negative).

\ifhpar \colorbox{colhd}{excessive pruning hypothesis} \\ \fi
Connections between neurons are created at a very high pace during postnatal development, to reach their maximum density around age six. Many connections are subsequently eliminated during the `pruning' phase which, for prefrontal cortex, peaks in late adolescence, a period that corresponds to the onset of most schizophrenia cases. A recent genetic study \citep{sekar2016} found a correlation between psychosis and a gene coding for a receptor that targets synapses for destruction by the immune system. According to the study's authors, the gene variant would cause excessive pruning and lead to schizophrenia. However, the defective gene is only present in a small minority of patients.

\ifhpar \colorbox{colhd}{role of traumas and structure} \\ \fi
A history of childhood traumas is reported to be a risk factor for schizophrenia. The content of positive symptoms is coherent with low self-esteem and negative affect, which are the common outcome of adverse childhood experiences \citep{schaefer2011}. The effect of traumas could be mediated and amplified by enhanced susceptibility \citep{horan2003, cohen2004}. The rest of the paper is organised as follows: the model is introduced in section 2 and 3, and used in section 4 to obtain an interpretation of some aspects of schizophrenia; section 5 draws the conclusions and outlines future research directions.

\section{A model for the mind}

\subsection*{Relational value}  

\ifhpar \colorbox{colhd}{features} \\ \fi
Reality can be conceived as a set of \textbf{situations}, each characterised through a list of active \textbf{features}. Examples of simple perceptual features are: `round shape', `horizontal orientation', `red color'. Examples of more abstract features are: `black ducks', `soccer players', `being a teacher'. Feature activation and deactivation is a continuous process, driven by perceptual features fed from sensory stimuli and propagated to more abstract ones in real time. This occurs on a fast time scale, as the mind `navigates' through everyday life. For instance, if a person is walking on the beach, the features encoding the concepts of `sand' and `sun' may be active, while features such as `office chair' and `computer screen' are likely to be inactive. 

\ifhpar \colorbox{colhd}{features, value} \\ \fi
We postulate that features are characterised by a property called \textbf{relational value}, that can be either positive or negative. Features such as `honesty' and `health', for example, are considered positive, while features such as `deception' and `illness' are generally perceived as negative. The determination and change of values happens by association: if a new feature with unknown value is presented in association with positive (negative) features, it will assume a positive (negative) connotation. Since a feature in general takes part in many associations, its value will be determined by the combined effect of all associations. The value is a long term property, expected to change on a slow time scale. 

\subsection*{A model for the mind}  

\begin{figure}[t] \begin{center} \hspace*{-0.50cm}
{\fboxrule=0.0mm\fboxsep=0mm\fbox{\includegraphics[width=18.00cm]{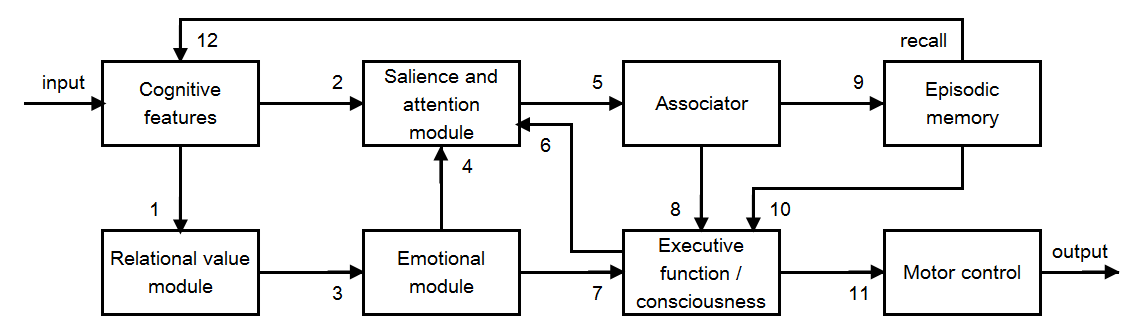}}}
\caption{Functional modules of the mind. The process of perception and memory formation starts with the activation of perceptual features triggered by external stimuli (input), proceeds with the addition of salience, continues with the generation of emotions, and ends with a new feature association presented to consciousness and stored in episodic memory. The same pathway is utilised when the record is recalled from memory and `played back' to consciousness. See text for details.}
\label{modules}
\end{center} \end{figure}

\ifhpar \colorbox{colhd}{modules /1} \\ \fi
The mind can be represented as a set of interacting functional modules (Fig.~\ref{modules}). \textbf{Cognitive features} are activated by the input flow and fed (link 1 in the figure) to the \textbf{relational value module} which adds value to active features and computes the value of two key symbols: self and object. This enriched information is passed (3) to the \textbf{emotional module}, that responds with an appropriate emotion. Features are also fed in parallel (2) to the \textbf{salience and attention module}, which uses the information about the current task pursued by the \textbf{executive module} (6) and the current emotional response (4) to produce features loaded with salience. 

\ifhpar \colorbox{colhd}{modules /2} \\ \fi
Salient features are passed to the associator (5), which binds them together and forms a new association. This association (8) and the current emotional response (7) are sent to the executive function, where consciousness is produced. The newly formed association in also stored in episodic memory (9). Finally, the executive function, based on present (7,8) and past (10) information, issues motor commands (11) and directs behaviour. The same pathway is triggered during the process of recall (12), by which an existing record is retrieved from memory, brought to consciousness and `played back'. 

\ifhpar \colorbox{colhd}{localisation /1} \\ \fi
Based on current neurobiological knowledge, the processing of perceptual and abstract features takes place in the cortex (e.g., occipital cortex for visual features). The relational and emotional modules could be localised in the medial prefrontal, cingulate and insular cortex, which are brain areas critical for social interaction \citep{palmiter2008, etkin2011, gu2013}.

\ifhpar \colorbox{colhd}{localisation /2} \\ \fi
The salience module could map to the striatum, a region involved in reward mechanisms \citep{msoelch2001}, in which dopamine is the dominant neurotransmitter. The associator corresponds to the hippocampus, a structure essential for memory formation. The executive function may be implemented in the prefrontal cortex, while episodic memory is distributed across the whole cortex.

\ifhpar \colorbox{colhd}{patient hm} \\ \fi
The independence of the consciousness generation pathway from the memory formation pathway is consistent with the findings of the research done on patient H.M. \citep{squire2009}, who had the hippocampus removed. Since then, he was unable to form new memories, but retained previously formed memories and was definitely capable of conscious awareness.   

\subsection*{Value computation, emotions and salience}  

\begin{figure}[t] \begin{center} \hspace*{-0.00cm}
{\fboxrule=0.0mm\fboxsep=0mm\fbox{\includegraphics[width=17.00cm]{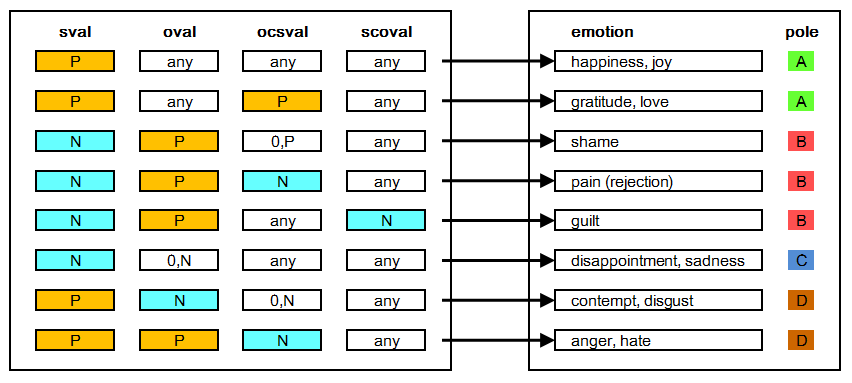}}}
\caption{Relational judgements and emotions. The relational module (on the left) adds value to features and computes the value of four variables (sval, oval, ocsval, scoval), that can be either positive (P), neutral (0) or negative (N). Based on the combination of these values, the emotional module (on the right) produces a specific emotion, belonging to one of four emotional poles.}
\label{emotpoles}
\end{center} \end{figure}

\ifhpar \colorbox{colhd}{relational module} \\ \fi
Social and hierarchical judgements play a central role in human behaviour. Emotions are often produced as a reaction to the actions of another person within a relational context, and depend on the appraisal of who is superior and who is inferior, who is right and who is wrong. The relational value module (Fig.~\ref{emotpoles}-left panel) assigns value to active features fed from the perceptual apparatus and calculates four variables: \textbf{sval} (self value), \textbf{oval} (object value), \textbf{ocsval} (object contribution to self value), \textbf{scoval} (self contribution to object value).  

\ifhpar \colorbox{colhd}{equations} \\ \fi
Self value $sval$ and object value $oval$ are in turn determined by the sum of the values of all \textit{active} features associated to them, $sfval_{i}$ and $ofval_{i}$. In formulas (S is the sigmoid function, which prevents values from exceeding a maximum threshold):

\begin{table}[h!]
\vskip 0.25cm
\center{
\begin{tabular}{C{8cm} C{8cm}}
$sval = S(\sum_{i}sfval_{i})$ & $oval = S(\sum_{i}ofval_{i})$
\end{tabular}}
\vskip 0.25cm
\end{table}

\noindent Since the set of active features changes depending on the situation, so does the self value. Therefore, in our model the self value (or self-esteem) is a spatio-temporally local concept: one person can have a high self-esteem in a given situation and a low self-esteem in another one. We can still define a global self-esteem as the mean self value across all situations.

\ifhpar \colorbox{colhd}{emotions} \\ \fi
Based on these variables, the emotional module (Fig.~\ref{emotpoles}-right panel) generates an emotion. Gratitude, for instance, is generated when sval and ocsval are both positive (in other words: when the self value is positive thanks to an action done by the object). Shame is produced when sval is negative and oval is positive (in other words: when the object is worthier than the self). Disappointment is elicited when sval is negative (in other words: moment characterised by low self-esteem). 

\ifhpar \colorbox{colhd}{emotional poles} \\ \fi
Emotions can be grouped in four \textbf{emotional poles}: \textbf{pole A} (happiness, love); \textbf{pole B} (shame, pain for rejection, guilt); \textbf{pole C} (disappointment, sadness); \textbf{pole D} (anger, contempt, envy). Pole A and pole D are characterised by positive self value and are therefore called \textbf{positive poles}, pole B and pole C are characterised by negative self value and are called \textbf{negative poles}.

\ifhpar \colorbox{colhd}{social status} \\ \fi
The association of shame, guilt and sadness to a lower social status is confirmed by numerous studies \citep{stevens1996}. Also the association of anger to a higher social status has been recognised and studied \citep{tiedens2011}. Other emotions, such as the fear of dying, appear to be more primordial and independent of social appraisal.  

\ifhpar \colorbox{colhd}{emotional salience} \\ \fi
Emotional experience is complemented by salience generation. Salience is a quality that defines the relative importance or prominence of (and attention paid to) a feature with respect to other features. In our model of brain architecture, salience is attached to features based on two criteria. The first criterion is the degree of contribution of the feature to the emotion generated (link 4 in Fig.~\ref{modules}). If, for instance, a person feels ashamed because he/she has bat ears, the feature `ears' receives high salience. In other words, the salience module is telling the mind: `you are not far from emotion E, which mostly depends on features X,Y: these are the features you need to monitor if you want to experience /avoid emotion E'.

\ifhpar \colorbox{colhd}{motivational salience} \\ \fi
The second criterion for salience attribution is the relevance of the feature for the task currently pursued by the executive function (link 6). If, for instance, a person is looking for the car key, all objects resembling keys get visual priority, while objects of different shapes are ignored. In this case, the message is: `features X,Y are linked to objective O on your agenda: these are the features you need to look at if you want to reach /avoid objective O'. The idea that neurons in the limbic system encode two kinds of salience (emotional and motivational) is not new \citep{bromberg2010}. 


\section{Normal and traumatic contexts}  

\subsection*{Normal functioning}  

\begin{figure}[t] \begin{center} \hspace*{-0.50cm}
{\fboxrule=0.0mm\fboxsep=0mm\fbox{\includegraphics[width=18.00cm]{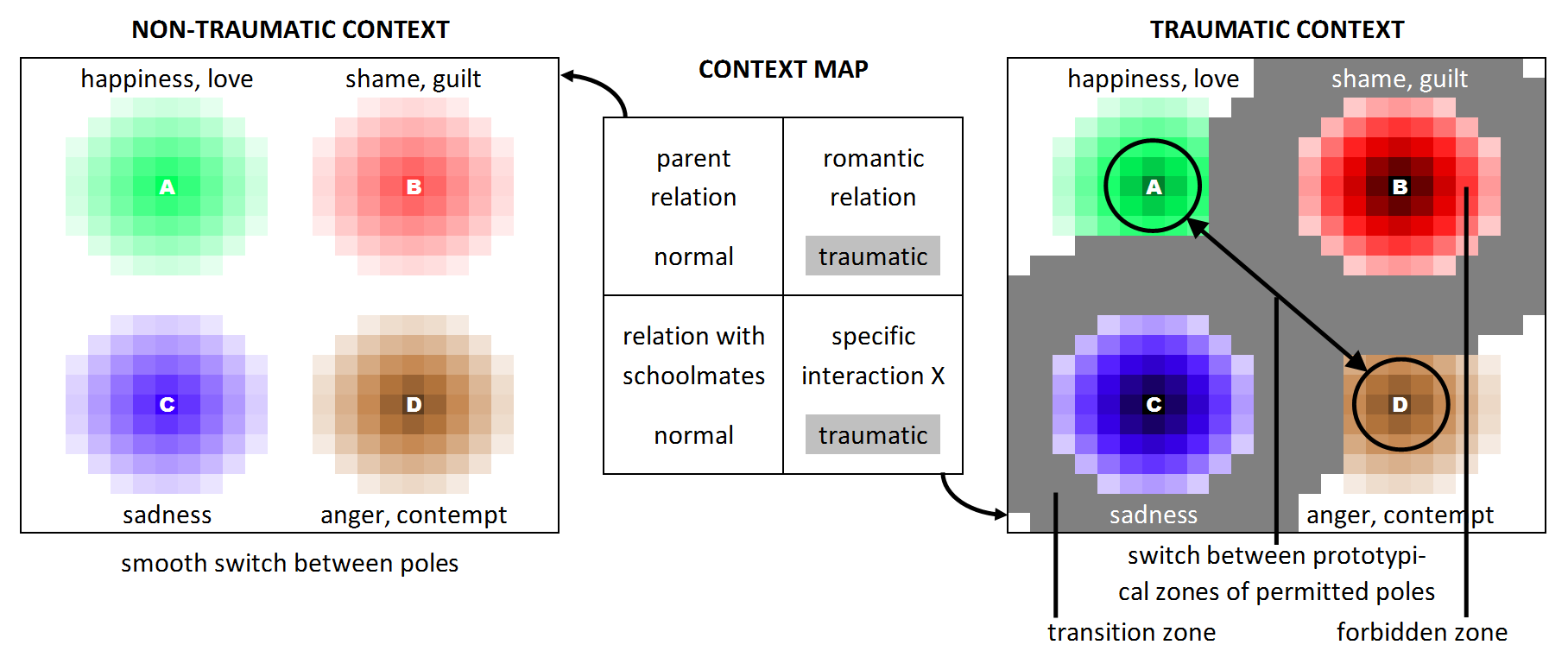}}}
\caption{Normal and traumatic contexts. The middle panel shows the map of all contexts, the left and right panels show the context plane for a non-traumatic and for a traumatic context respectively. Each point in the context plane represent a situation (defined as sets of coativated features) and is associated to an emotion (whose intensity is represented by the colour shade). In a normal context (left panel), the mind can switch smoothly between emotional poles, no pole requires dissociation and the repertoire of emotions is fully accessible. In a traumatic context (right panel), pole B and/or pole C are characterised by too intense emotional levels, and are inhibited. When the mind happens to be in one of these poles (forbidden zone), dissociation intervenes. To avoid dissociation, the mind oscillates between poles A and D, staying in each pole as long as the situation is prototypical for the pole. When the situation drifts out of the free zone into the transition zone (shown in grey), splitting symptoms appear, and the mind switches to the free zone of the other permitted pole.}
\label{switchx}
\end{center} \end{figure}

\ifhpar \colorbox{colhd}{contexts and situations} \\ \fi
Let us define a \textbf{situation} as a set of features simultaneously active. Examples of situations are: `piano lesson with uncle' (coactivated features: image of piano, image of hands on the keyboard, sound of uncle's voice, sound of piano, etc.); `tennis match with a friend' (coactivated features: image of racquet, image of opponent, sound of racquet hitting the ball, odour of sweat, etc.). Situations can be thought of as belonging to different \textbf{contexts}, such as `parent relation', `romantic relation', `relation with schoolmates', etc. (Fig.~\ref{switchx}-mid panel). A context can be conveniently represented on a plane, in which points correspond to individual situations (Fig.~\ref{switchx}-left panel). 

\ifhpar \colorbox{colhd}{poles on planes} \\ \fi
Each point is associated to an emotion and belongs to the `zone of influence' of one of the four emotional poles. Each pole originates from a point representing the most prototypical situation associated to the pole, and extends towards less prototypical situations. The epicentre of Pole D, for instance, may correspond to a situation-point characterised by an object behaving very dishonestly, eliciting a very strong anger, while points further away may be characterised by a better behaviour of the object (a pole `hosts' more emotions and therefore in general has more `centres': only one centre is shown in the figure for simplicity). We define a context `normal' if the highest emotional levels are not too high. In this condition the mind can switch smoothly between all poles, experiencing different levels of the emotions associated to each pole (Fig.~\ref{switchx}-left panel).

\subsection*{Traumatic functioning: dissociation}

\ifhpar \colorbox{colhd}{conditions for trauma} \\ \fi
The emotional poles that can be involved in a trauma are the negative poles B and C. In our model, a trauma occurs when the intensity of the elicited emotions is too high: in case of pole B, based on what we said in section 2, this requires the existence of an value gap between object and self (object worthier than self). This in turn requires that object-associated features have on average a higher value than self-associated features. If a person with bat ears, big nose and thin lips thinks that these features are very negative and he/she gets criticised or made fun of for them, a trauma may take place.

\ifhpar \colorbox{colhd}{dissociation} \\ \fi
Upon the occurrence of a trauma, the associator stops working: this corresponds to the phenomenon of \textbf{dissociation}, defined as the distortion, limitation or loss of the normal associative links between perceptions, emotions, thoughts and behaviour. Dissociation can take the form of mental `black-out', depersonalisation (feeling of separation from one's body), derealisation (feeling of being detached from the world), selective amnesia and emotional detachment \citep{lanius2015, radovic2002depers}.   

\subsection*{Traumatic functioning: splitting}

\ifhpar \colorbox{colhd}{dissociation cost} \\ \fi
In case of trauma, the adoption of dissociation makes it possible for the mind the stay on negative poles, excluding the awareness of intolerable thoughts and emotions from consciousness. However, the disconnection of aspects of reality may hide potential dangers and have a high cost. Therefore, the mind tries to avoid the negative poles B and C, and heads towards the positive poles A and D, where the perception of reality is not restricted. The space of a traumatic context can be divided into three zones (Fig.~\ref{switchx}-right): the `forbidden zone', an area near a traumatic pole which cannot be accessed in a non-dissociated state; the `free zone', an area far from all traumatic poles (or near the positive poles A and D); the `transition zone', a safety belt around the forbidden zone.  

\ifhpar \colorbox{colhd}{switch cycle} \\ \fi
Let us assume that, in a traumatic context, the mind is initially near pole A. The mind will stay in the free zone around pole A as long as conditions are prototypical for pole A, i.e. as long as the object relation is perfect, full of trust, mutual respect, etc. As the situation departs from the prototypical scenario of pole A and drifts into the transition zone, the mind switches abruptly to the free zone around pole D. When the situation deviates from the prototypical scenario of pole D, the mind returns to pole A, and the cycle repeats itself. This corresponds to the defense mechanism of \textbf{splitting}, defined as the inability to integrate positive and negative aspects of self and others, which results is a view of the world in `black and white' \citep{perry2013}.

\ifhpar \colorbox{colhd}{transition zone, salience} \\ \fi
The transition zone is a safety belt built around the forbidden zone, characterised by high emotional levels. This causes an increase of the emotional salience of the features involved in the trauma (link 4 in Fig.~\ref{modules}), which represent the person's defects criticised. As a result, these features become the focus of attention and appear magnified and distorted: if the feature criticised is `bat ears', when looking at the mirror the person will see his/her ears magnified and more protruded, like in a caricature. The magnification of defects serves the purpose to warn the executive function that the current situation is close to a traumatic point, and gives an indication of the features that need to be closely monitored. 

\ifhpar \colorbox{colhd}{pseudo-hallucinations in various disorders} \\ \fi
This phenomenon is reminiscent of symptoms relevant to a wide range of mental disorders, starting from the hallucinations, the delusions and, more in general, the `delusional atmosphere' that characterise schizophrenia \citep{moskowitz2008}. Transitory perceptual distortions or `pseudo-hallucinations' are not uncommon also in personality disorders \citep{gras2014}. Perceptual distortions are present in body dysmorphic disorder, which can either appear stand-alone \citep{phillips2004} or be responsible for the dysmorphic body image associated to eating disorders (among many other conditions) \citep{ruffolo2006}. 

\subsection*{Selective dissociation}

\begin{figure}[t] \begin{center} \hspace*{-0.0cm}
{\fboxrule=0.0mm\fboxsep=0mm\fbox{\includegraphics[width=14.00cm]{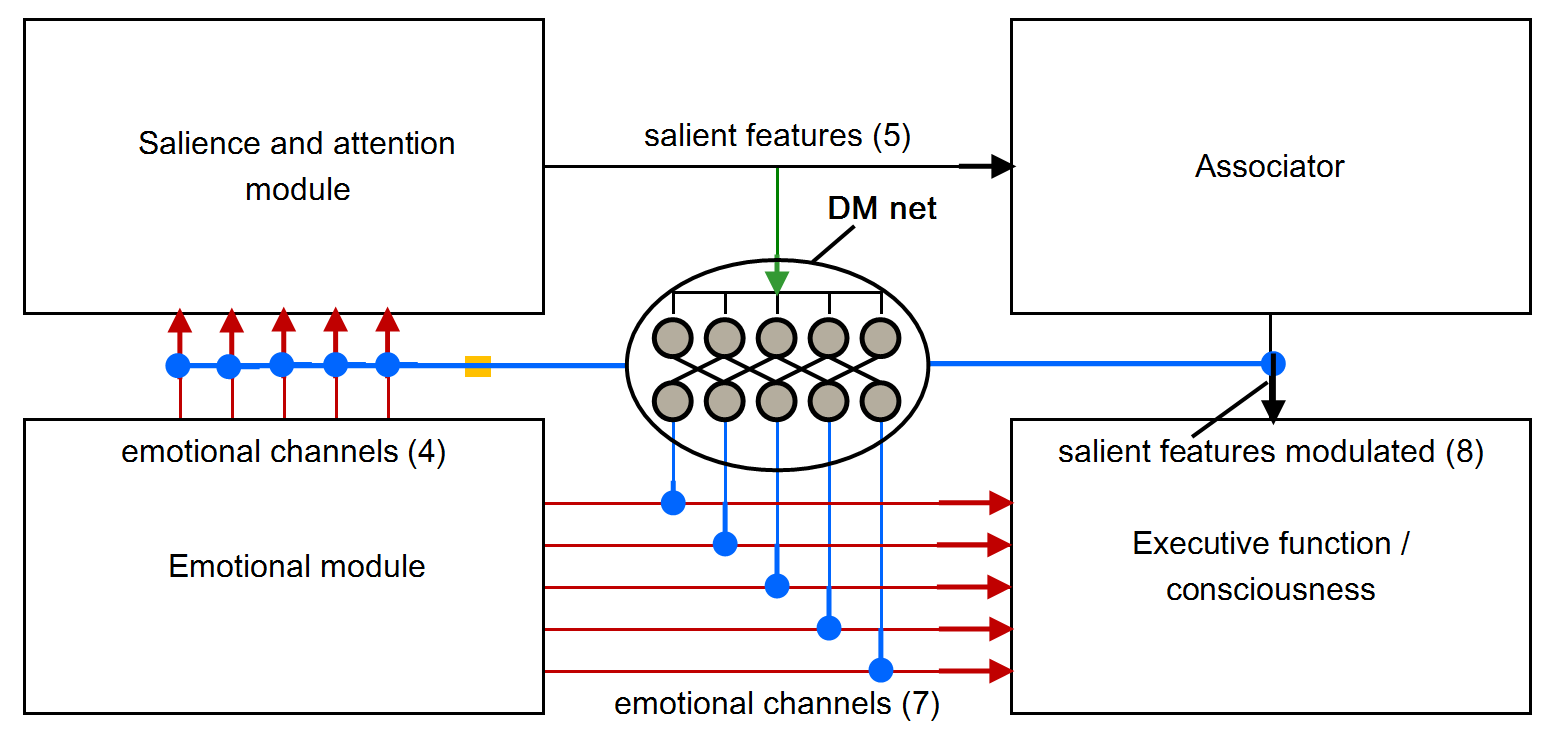}}}
\caption{Emotional dissociation, dissociative modulatory network. This figure shows a magnification of some of the modules in Fig.~\ref{modules}. The DM network receives in input cognitive features loaded with salience (green arrow) and sends modulatory signals (shown in blue) to emotional channels and to cognitive features (4, 7, 8). The effect of modulation is perceived by the executive module as a restriction of consciousness. Cognitive features are also sent to the associator (5), which binds them together and stores the new association in episodic memory. The memory record is complete, including the dissociated components: once the record is recalled from memory, such components are again dissociated and excluded from consciousness.}
\label{dmnet}
\end{center} \end{figure}

\ifhpar \colorbox{colhd}{first deep, then selective dissociation} \\ \fi
When the mind leaves the transition zone and enters the forbidden zone, dissociation is used to avoid emotional pain. We can hypothesise that, upon the first occurrence of a trauma, dissociation is total and involves all mental functions: this may correspond to the `freezing' behaviour of `disorganised attached' children \citep{main1986}. Total dissociation provides a shield to pain, but may expose the individual to serious consequences in a potentially dangerous environment: for this reason, it cannot be sustained for long periods of time. It is not unrealistic to assume that, in case of repeated traumas, the mind will try to replace total dissociation through less pervasive forms of dissociation, able to preserve a reasonable level of functioning.

\ifhpar \colorbox{colhd}{selective dissociation, examples} \\ \fi
\textbf{Selective dissociation} is a limited form of dissociation which, near a traumatic situation, excludes from consciousness \textit{some} emotional or cognitive channels. This is a common phenomenon. Medical students may be shocked when they see an operation for the first time, but they rapidly get used to it. The same happens to the medical staff of intensive care units, where death is a daily occurrence, or to the employees of a slaughterhouse. This happens to all of us, when we see poor people living in the street and pretend it is normal. We can think of this process as of a form of learning, in which the mind selects the smallest subset of reality that needs to be dissociated to avoid emotional pain without losing touch with the `here and now'.      

\begin{figure}[t] \begin{center} \hspace*{-0.0cm}
{\fboxrule=0.0mm\fboxsep=0mm\fbox{\includegraphics[width=14.00cm]{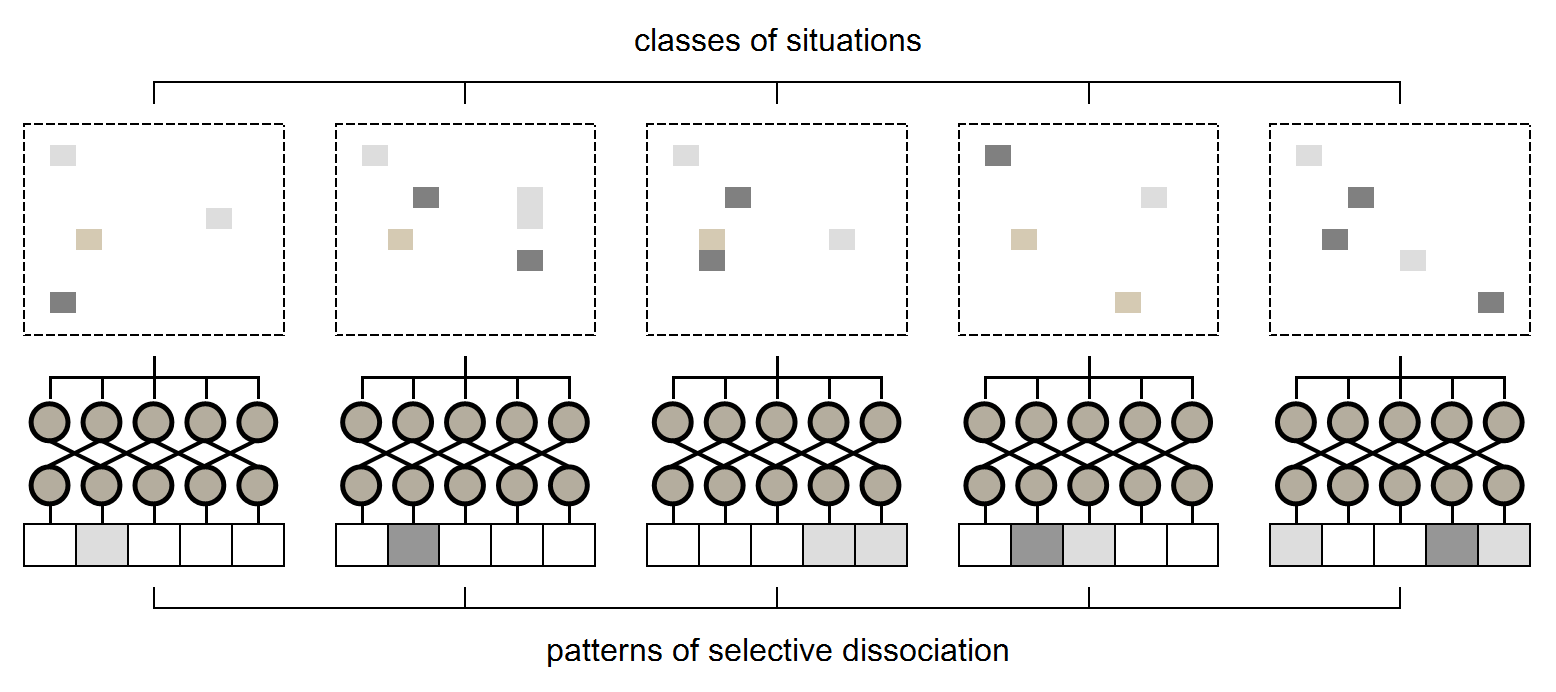}}}
\caption{Classification of situations and patterns of selective dissociation. The DM network classifies the input composed of cognitive features into a number of classes. For each class, it produces a specific pattern of selective dissociation, i.e. a combination of attenuation levels for cognitive and emotional channels. Depending on the number of available neurons, the DM network can learn to map a different number of classes.}
\label{dissocpat}
\end{center} \end{figure}

\ifhpar \colorbox{colhd}{DM network} \\ \fi
The learning of selective dissociation can be obtained through a modulation of links 4, 7 and 8 of Fig.~\ref{modules}, which could be attenuated in certain situations. We hypothesise that this is achieved through a dedicated \textbf{dissociative modulatory (DM) network} that receives in input cognitive features loaded with salience and sends modulatory signals to emotional and cognitive links (Fig.~\ref{dmnet}). In other words, for each situation, this network decides the degree of attenuation of each channel. As a result, different classes of situations are selectively mapped to different attenuation patterns (Fig.~\ref{dissocpat}). The modulated signals are subsequently brought to consciousness, which is restricted thanks to the effect of modulation. 

\ifhpar \colorbox{colhd}{complete memory records} \\ \fi
On the other hand, there is ample evidence for the existence of subconscious, dissociated memory records, which are denied access to consciousness, but which can re-emerge after a period of time, when the conditions that justified their suppression have been removed. Painful memories, such as incest or domestic violence, can be reactivated after many years, when the person feels safe (e.g. in a therapeutic setting). For these reason, we hypothesise that the records stored in memory (through link 10) are complete, including the dissociated components. Once the record is recalled from memory (12), such components are again dissociated and excluded from consciousness, until the DM network configuration changes. 

\ifhpar \colorbox{colhd}{emotions on-the-fly} \\ \fi
We further hypothesise that memory records are only composed of cognitive features. Emotions are not recorded in memory, but generated `on-the-fly' based on the set of active features, once features are activated, through either external input or memory recall. This explains how the emotions associated to an event can change with time (e.g., painful memories become less painful as a result of therapy) even though the cognitive memory of the event is unchanged.


\section{Interpretation of schizophrenia}

\ifhpar \colorbox{colhd}{traumatic context} \\ \fi
Childhood traumas involving caregivers are often present in the history of persons with schizophrenia \citep{schaefer2011}. As a result, the mental functioning may be altered in key contexts such as `parent relation' and `romantic relation' (Fig.~\ref{switchx}). In such contexts poles B and C are difficult to access, and the mind is forced to oscillate between poles A and D (splitting). In the transition zone, the features involved in the trauma are overloaded with salience and appear magnified and distorted, like in a caricature: we call such distortions `pseudo-hallucinations'. When the mind gets too close to a traumatic point, selective dissociation should intervene on specific cognitive and emotional channels, removing disturbing thoughts and turning down the `volume' of bad emotions. 

\ifhpar \colorbox{colhd}{weak dissociation, aberrant salience} \\ \fi
Our hypothesis is that the learning process that leads to selective dissociation is defective in schizophrenia: dissociation is either too weak or too strong. When it is too weak, painful emotions are not attenuated, leaving the mind unprotected near the epicentre of the trauma. As a result, feature distortion caused by excessive salience proceeds unabated and becomes more pronounced: this is, in our opinion, the nature of hallucinations. 

\ifhpar \colorbox{colhd}{weak dissociation, unreliable associations} \\ \fi
When the self and its associated features are strongly devalued, they are `responsible' for the shame, guilt or pain experienced (they are the major contributors to the emotion): as such, they are heavily loaded with salience. In practical terms, this means that such features are `presented' as important elements, and end up dominating the conscious attention of the subject. In such conditions, the disabling of the associator (dissociation) is necessary to avoid the establishment of unreliable associations, lacking statistical robustness. If dissociation does not intervene in a timely manner, salient features, such as the self, run the risk of being associated with random features that happen to co-occur. This explains the self-referential nature of delusions in schizophrenia.

\ifhpar \colorbox{colhd}{strong dissociation, negative symptoms} \\ \fi
Dissociation eventually intervenes, too strongly. When the dissociation wave hits the brain, it produces the full spectrum of negative symptoms: emotional detachment, anhedonia, lack of motivation and thought disorganisation. The degree of dissociation correlates with the severity of the symptoms observed. Even if the DM network is defective, some learning may still be expected, leading to a limited progressive improvement over time. 

\ifhpar \colorbox{colhd}{dopamine} \\ \fi
The dopamine hypothesis maintains that hallucinations and delusions are caused by an excess of salience, in turn driven by an excess of dopamine in the limbic system. Most antipsychotic drugs block the dopamine receptor: this lowers the level of dopamine, reduces `aberrant' salience and eliminates the positive symptoms. However, based on our model, the excess of dopamine is not due to a malfunction of the salience module, but it is caused by too intense emotional levels, left undissociated by a defective DM network. An intervention on dopamine blocks the salience mechanism in all situations and lead to an abnormal functioning of the brain, which may exacerbate the negative symptoms.     

\begin{figure}[t] \begin{center} \hspace*{-0.0cm}
{\fboxrule=0.0mm\fboxsep=0mm\fbox{\includegraphics[width=12.00cm]{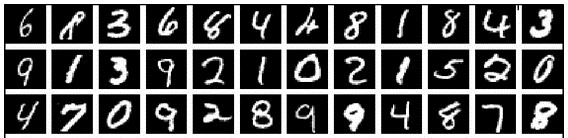}}}
\caption{Classification of handwritten digits. The Mnist dataset, used in machine learning for visual classification tasks, contains 60000 examples of handwritten characters belonging to 10 classes (0-9).}
\label{mnist}
\end{center} \end{figure}

\ifhpar \colorbox{colhd}{classification task} \\ \fi
It is not unreasonable to think that the classification power of the DM network (the number of classes that can be discriminated) depends on the number of neurons in the network. A typical example of classification task in the field of artificial neural networks is the classification of handwritten digits (see Fig.~\ref{mnist}). In this case, the neural network receives in input the image of the character and produces in output an integer number in the [0,9] range representing the class. 

\ifhpar \colorbox{colhd}{classification power} \\ \fi
In order to effectively solve a given classification task, an artificial neural network must have a sufficient number of neurons. If the number of neurons is too low, the classification will be inaccurate (e.g. a `2' will be misclassfied as a `4'). The number of neurons required depends on the complexity of the task, which in turn depends on the dimensionality of the input (the number of pixels in the case of images) and the dimensionality of the output (the number of classes). 

\ifhpar \colorbox{colhd}{poor classification} \\ \fi
Our conjecture is that, in schizophrenia, the DM network has an insufficient number of neurons. As a result, the classification ability of the DM network is poor. Instead of providing a customised dissociational pattern for each traumatic point, the DM network is only able to generate a few distinct patterns. It is as though we tried to fit all feet sizes with only two shoe measures, 38 and 45: for some feet, the shoe will be too large, for other feet it will be too small (for very few, it will be ok). Likewise, for some situations the dissociational pattern will be too weak, producing positive symptoms, for other situations it will be too strong, producing negative symptoms.  
 
\ifhpar \colorbox{colhd}{defective hippocampus} \\ \fi
The hypothesised reduced number of neurons in the DM network correlates with the observation of structural changes to the hippocampus frequently observed in schizophrenia \citep{heckers2010} (it is reasonable to assume that the DM network is located near the hippocampus). The hippocampus of schizophrenic subjects is significantly smaller and seems to be less susceptible to the phenomenon of habituation \citep{williams2013}, which could be necessary to develop selective dissociation. These observations led to the hypothesis of psychosis as a learning and memory problem \citep{tamminga2013}, coherent with our theory.

\ifhpar \colorbox{colhd}{non-trauma-linked factors} \\ \fi
Several environmental non-trauma-linked factors, such as difficulties during birth, or the use of drugs, are correlated with schizophrenia. The level of heritability is high: if a twin has schizophrenia, the probability that also the other twin has it is 50\%. Overall, there is evidence that, besides traumatic experiences, biological, genetic or neurodevelopmental factors play a role in the pathogenesis of schizophrenia: our hypothesis is that all these factors act by reducing the number of neurons in the DM network. This represents a predisposition that requires the action of traumas to be realised. The accumulation rate of traumatic events is expected to be constant in the course of life and to accelerate during adolescence, which is acknowledged to be a stressful period. This would explain why schizophrenia has its onset during adolescence. 

\section{Conclusions}

\ifhpar \colorbox{colhd}{xxxx} \\ \fi
The objective of this work was to propose a functional model of the mind and provide a new interpretation for schizophrenia. The model evidences the interplay between environmental causes and neurobiological predisposition in the generation of the schizophrenic phenotype. The environmental contribution consists in the presence of significant traumas, while the intrinsic susceptibility depends on the inability of the DM network to generate a sufficient number of customised dissociation patterns, to counteract the effect of traumas in all situations. Future work will be aimed to further develop the model and draw from it insights useful for therapy. 


\bibliographystyle{apalike}
\bibliography{lschizophr} 

\end{document}